%Paper: gr-qc/9303030
%From: snh@ibm-1.MPA-Garching.MPG.DE (Sean Hayward)
%Date: Fri, 26 Mar 93 15:55:19 +0100

\font\lbf=cmbx10 scaled\magstep2
\font\sm=cmr7

\def\bs{\bigskip}
\def\ms{\medskip}
\def\np{\vfill\eject}
\def\nl{\hfill\break}
\def\ni{\noindent}
\def\cl{\centerline}

\def\title#1{\cl{\lbf #1}\ms}
\def\ctitle#1{\bs\cl{\bf #1}\par\nobreak\ms}

\def\ref#1#2#3#4{#1\ {\it#2\ }{\bf#3\ }#4\par}
\def\ns{\kern-.33333em}
\def\ANY{Ann.\ NY Acad.\ Sci.}
\def\CQG{Class.\ Qu.\ Grav.}
\def\CMP{Comm.\ Math.\ Phys.}
\def\CPAM{Comm.\ Pure \& App.\ Math.}
\def\GRG{Gen.\ Rel.\ Grav.}
\def\JMP{J.\ Math.\ Phys.}
\def\JP{J.\ Phys.}
\def\PL{Phys.\ Lett.}
\def\PR{Phys.\ Rev.}
\def\PRL{Phys.\ Rev.\ Lett.}
\def\PRS{Proc.\ R.\ Soc.\ Lond.}

\def\address{\ms\cl{Max-Planck-Institut f\"ur Astrophysik}
		\cl{Karl-Schwarzschild-Stra\ss e 1}
		\cl{8046 Garching bei M\"unchen}
		\cl{Germany}\ms}
\def\me{\ms\cl{\bf Sean A. Hayward}\address}

\def\eg{{\it e.g.\ }}
\def\ie{{\it i.e.\ }}
\def\cf{{\it cf\/\ }}
\def\etc{{\it etc}}
\def\etal{{\it et al\/\ }}

\magnification=\magstep1

\def\H{{\cal H}}
\def\L{{\cal L}}
\def\R{{\cal R}}
\def\k{\kappa}
\def\t{\theta}
\def\s{\sigma}
\def\o{\omega}
\def\N{\nabla}
\def\p{\partial}
\def\half{{\textstyle{1\over2}}}
\def\root{{\textstyle{1\over{\sqrt2}}}}
\def\ADM{{\sm ADM}\ }
\def\Bondi{Bondi \etal}
\def\tt{{\tilde\t}}
\def\st{{\tilde\s}}
\def\nt{{\tilde\nu}}
\def\sss{\s_{ab}\s^{ab}}
\def\sst{\s_{ab}\st^{ab}}
\def\oo{\o_a\o^a}
\def\I{\int_S}
\def\A{\sqrt{{A\over{16\pi}}}}
\def\tv{\vartheta}
\def\pv{\varphi}
\def\l{\lim_{r\to\infty}}
\def\P{\Psi_2}
\def\Sch{\left(1-{2m\over r}\right)}
\def\RN{\left(1-{2m\over r}+{e^2\over{r^2}}\right)}
\def\sph{(d\tv^2+\sin^2\tv\,d\pv^2)}
\def\a{\sin\tv\,d\tv\wedge d\pv}
\def\r{{\hat r}}

\def\C{h^{ac}h^{bd}C_{abcd}}
\def\m{{\bar m}}
\def\Ph{{\hat\P}}
\def\sh{{\hat\s}}
\def\sb{{\hat{\bar\s}}}
\def\lh{\lim_{\r\to\infty}}
\def\Ah{\sqrt{{{\hat A}\over{16\pi}}}}
\def\Sh{{\hat S}}
\def\Ih{\int_\Sh}
\def\mh{{\hat\mu}}

\title{Quasi-local gravitational energy}
\me
\cl{Revised 16th March 1993}
\ms
{\bf Abstract.}
A dynamically preferred quasi-local definition of gravitational energy
is given in terms of the Hamiltonian
of a `2+2' formulation of general relativity.
The energy is well-defined for any compact orientable spatial 2-surface,
and depends on the fundamental forms only.
The energy is zero for any surface in flat spacetime,
and reduces to the Hawking mass in the absence of shear and twist.
For asymptotically flat spacetimes,
the energy tends to the Bondi mass at null infinity
and the \ADM mass at spatial infinity,
taking the limit along a foliation parametrised by area radius.
The energy is calculated for the Schwarzschild, Reissner-Nordstr\"om
and Robertson-Walker solutions, and for plane waves and colliding plane waves.
Energy inequalities are discussed,
and for static black holes the irreducible mass is obtained on the horizon.
Criteria for an adequate definition of quasi-local energy are discussed.

\ctitle{1. Introduction}
It comes as a surprise to many
that there is no agreed definition
of gravitational energy (or mass) in general relativity.
In Newtonian gravity,
there is a material density which may be integrated over a spatial 3-surface
to give the mass,
and the Poisson equation and Gauss theorem then yield
an expression for the mass inside a 2-surface,
measured on the 2-surface---a so-called quasi-local mass.
Finding a similar quasi-local expression
for the Newtonian gravitational energy is impossible,
since the energy density, unlike the material density,
is not a total divergence.
In general relativity, material mass is merely one aspect of
the energy-momentum-stress tensor, and gravitational energy,
if well-defined at all, is non-local,
as follows from the equivalence principle.
The gravitational field can be measured by
the geodesic deviation of two observers,
but a single observer cannot distinguish it from kinematical effects.
Equivalently, curvature cannot be measured on a point or line,
but requires a 2-surface at least.
\ms
In the case of an asymptotically flat spacetime,
the Bondi and \ADM masses are well-defined asymptotic quantities
which are generally accepted as the total mass of the spacetime
measured on spheres at null infinity and spatial infinity respectively
(\Bondi 1962, Sachs 1962, Arnowitt \etal 1962).
The existence of these asymptotic definitions,
and in particular the Bondi mass-loss result,
indicate that gravitational energy is not meaningless,
and that there is a real physical process which is responsible for
radiating gravitational energy to infinity in such a spacetime.
Since an appropriate definition of this energy cannot be found locally,
a quasi-local definition is sought,
where `quasi-local' is here taken in the conventional sense of referring to
a compact orientable spatial 2-surface, usually spherical.
This is the natural finite analogue of the asymptotic definitions.
\ms
Hawking (1968) defined a quasi-local mass which has various desirable
properties: it is zero for a metric sphere in flat spacetime,
it gives the correct mass for the Schwarzschild solution
on a metric sphere, it tends to the Bondi mass asymptotically
in a static asymptotically flat spacetime,
and it tends to zero as a sphere in any spacetime is shrunk to a point
(Eardley 1979, Horowitz and Schmidt 1982).
Unfortunately, it is non-zero for generic 2-surfaces in flat spacetime.
This drawback is easily corrected by adding a certain term,
as is shown subsequently.
\ms
Penrose (1982) emphasised the conceptual importance of quasi-local mass,
and suggested a twistorial construction.
This can be properly defined if the 2-surface can be transplanted into
a conformally flat spacetime (Tod 1983, 1986),
but the transplant cannot be done generically without damaging
the structure of the 2-surface (Helfer 1992).
Bergqvist (1992a) reviews various other attempts,
of which that of Dougan and Mason (1991) comes closest to being
a well-defined functional of 2-surfaces, being well-defined generically,
but breaking down for the important case of marginally trapped 2-surfaces.
Bergqvist also shows that seven different definitions
all give different results in two simple examples,
namely the Reissner-Nordstr\"om solution and the Kerr horizon.
This is an unfortunate situation which indicates the need
for a more critical approach to what constitutes an adequate definition.
Currently, there seems to be no demand for a quasi-local energy
to be even well-defined, let alone geometrically or dynamically natural.
\ms
The main purpose of this article is to present, in \S3,
a dynamically preferred quasi-local energy which is essentially
the `2+2' Hamiltonian of the Einstein gravitational field.
This is a well-defined functional of compact orientable spatial 2-surfaces,
and is geometrically natural in the sense of depending only on
the fundamental forms of the 2-surface, introduced in \S2.
The energy is shown to vanish for any surface in flat spacetime in \S4,
to tend to the Bondi mass at null infinity in \S5,
and to tend to the \ADM mass at spatial infinity in \S6,
if the spacetime is asymptotically flat.
Uniqueness on purely geometrical grounds is considered in \S7,
examples in \S8, energy inequalities in \S9,
and criteria for an adequate definition in \S10.

\ctitle{2. Geometry of 2-surfaces}
It is appropriate to begin with a review of the geometry of
a compact orientable spatial 2-surface $S$ embedded in spacetime,
according to the `2+2' formalism developed by the author (Hayward 1993a),
which describes null foliations of such surfaces.
In this formalism, a basis $(e^a_1,e^a_2)$ for $S$ is completed to a
spacetime basis $(u^a,v^a,e^a_1,e^a_2)$ by Lie propagation:
$(\L_u,\L_v)(u^a,v^a,e^a_1,e^a_2)=0$, where $\L$ denotes the Lie derivative.
The commuting vectors $(u^a,v^a)$ are referred to as the evolution vectors,
since Lie propagation in these directions enables the spacetime
in a neighbourhood of $S$ to be developed.
(This is analogous to the Cauchy problem, where the single evolution direction
may be decomposed into a lapse function and a shift vector.)
The choice of evolution vectors is partially fixed by demanding that
the 3-surfaces $(\L_uS,\L_vS)$ are null,
$u^al_a=v^an_a=0$,
where the null normals $(l_a,n_a)$ satisfy
$$\eqalign{
\N_{[a}l_{b]}=\N_{[a}n_{b]}=0,\qquad
&l_al^a=n_an^a=0,\cr
l_an^a=-e^m,\qquad
h_{ab}l^b=h_{ab}n^b=0,\qquad
&g_{ab}=h_{ab}-e^{-m}(l_an_b+n_al_b),\cr}
$$
and $g_{ab}$ is the spacetime metric, $h_{ab}$ the induced 2-metric of $S$
and $m$ is called the scaling function.
The remaining freedom in the evolution vectors is given by the shift 2-vectors
$r^a=h^a_bu^b$
and $s^a=h^a_bv^b$.
The metric then decomposes into
$$
g_{ab}=\pmatrix{
r_cr^c&r_cs^c-e^{-m}&r_b\cr
r_cs^c-e^{-m}&s_cs^c&s_b\cr
r_a&s_a&h_{ab}\cr}
$$
in the above basis.
The freedom to interchange or boost the null normals is left open.
\ms
The dynamically independent first derivatives of the metric
consist of the expansions
$$
\t=\half h^{cd}\L_{u-r}h_{cd},\qquad
\tt=\half h^{cd}\L_{v-s}h_{cd},
$$
the (traceless) shears
$$
\s_{ab}=h_a^ch_b^d\L_{u-r}h_{cd}-\half h_{ab}h^{cd}\L_{u-r}h_{cd},\qquad
\st_{ab}=h_a^ch_b^d\L_{v-s}h_{cd}-\half h_{ab}h^{cd}\L_{v-s}h_{cd},
$$
the `inaffinities'
$$
\nu=\L_{u-r}m,\qquad
\nt=\L_{v-s}m,
$$
and the `twist', or `anholonomicity', or commutator of the null normals,
$$
\o_a=\half e^mh_{ab}(\L_us^b-\L_vr^b-\L_rs^b).
$$
These fields encode the extrinsic curvature of $S$,
or equivalently the momenta conjugate to the configuration fields
$(h_{ab},r^a,s^a,m)$.
In Hayward (1993a), the vacuum Einstein equations are written
in a first-order form in terms of
$(h_{ab},r^a,s^a,m,\t,\tt,\s_{ab},\st_{ab},\nu,\nt,\o_a)$,
which constitutes a `2+2' analogue of the \ADM `3+1' formalism.
The initial data for the associated characteristic initial value problem
can be taken as  $(h_{ab},m,\t,\tt,\o_a)$ on $S$,
$(\s_{ab},\nu)$ on $\L_uS$ and
$(\st_{ab},\nt,s^a)$ on $\L_vS$,
with $r^a$ being prescribed over the whole 4-d patch.
Of this data, $r^a$, $s^a$, $m$, $\nu$ and $\nt$ represent coordinate freedom
on the respective surfaces, and will be set to zero henceforth.
\ms
The remaining dynamical variables are, in traditional language,
the first fundamental form $h_{ab}$,
the (null) second fundamental forms $\s_{ab}+\t h_{ab}$
and $\st_{ab}+\tt h_{ab}$,
and the normal fundamental form $\o_a$ (\cf Spivak 1979),
with the null normals fixing a preferred basis
for the fibres of the normal bundle, up to boosts and interchange.
These fundamental forms are the only geometrical invariants of a 2-surface
in spacetime, and are also the dynamically independent parts of the
gravitational field referred to a 2-surface.
It is therefore most natural to seek a definition of quasi-local energy
in terms of these forms. In fact, of the various definitions,
this is the case only for the Hawking mass,
$$
M_H={1\over{8\pi}}\A\I\mu(\R+\t\tt),\qquad
A=\I\mu,
$$
where $\mu$ is the area 2-form and $\R$ the Ricci scalar of $h_{ab}$.

\ctitle{3. Definition of the energy}
To measure the energy of a field on a spatial 3-surface,
the Hamiltonian of a `3+1' formulation is usually taken.
On a compact orientable spatial 2-surface $S$,
the analogous dynamical object for the Einstein field
is the Hamiltonian 2-form,
which may be quoted from Hayward (1993a) as
$$
8\pi\H=-\mu(\R+\t\tt-\half\sst-2\oo),
$$
where the notation and choice of evolution are as in \S2,
and the $8\pi$ has been inserted to agree with the units $G=1$
used in most papers on quasi-local energy.
This is a preferred dynamical quantity in the sense that variation
of the (full) Hamiltonian yields the Einstein equations
in a form adapted to null foliations of spatial 2-surfaces.
The Hamiltonian is obtained from a Lagrangian which is
the Einstein-Hilbert Lagrangian in `2+2' form, up to a total divergence.
In a `3+1' context, the energy is defined as the integral of
the Hamiltonian over the 3-surface,
but in a `2+2' context, $\I\H$ needs to be multiplied by a length
to give energy units, and the only natural length scale is given by the area
$$
A=\I\mu.
$$
The proposed definition of the gravitational energy of a 2-surface,
or its quasi-local energy, is then
$$
E=-\A\I\H,
$$
where the factor has been chosen for agreement with the Schwarzschild mass.
Thus
$$
E={1\over{8\pi}}\A\I\mu(\R+\t\tt-\half\sst-2\oo).
$$
It should be noted that the expression is valid for any one 2-surface,
but that when considering the variation of energy between surfaces,
the quasi-local coordinate freedom cannot be fixed as in \S2
and the full Hamiltonian
%$$\eqalign{
%8\pi\H&=
%-\mu(\t\tt-\tt\nu-\t\nt-\half\sst)-\mu e^{-m}(\R-2\oo+\half\D_am\D^am)\cr
%&\qquad+\half\mu(\st^{ab}+(\nt-\tt)h^{ab})\L_rh_{ab}
%+\half\mu(\s^{ab}+(\nu-\t)h^{ab})\L_sh_{ab}\cr
%&\qquad+\mu\tt\L_rm+\mu\t\L_sm+2\mu\o_a\L_rs^a\cr}
%$$
should be used.
%where $\D_a$ is the covariant derivative operator of $h_{ab}$.
%Although rather lengthy, this is at least an explicit global definition
%which can be evaluated directly.
%It simplifies dramatically for a null 3-surface,
%where $r^a$, $s^a$, $m$ and either $\nu$ or $\nt$ can be set to zero
%by coordinate choice.
\ms
Various immediate observations concerning $E$ can be made. Firstly,
the total energy of two surfaces $S=S_1\cup S_2$
is greater than the sum of the individual energies, since
$$
A=A_1+A_2,\qquad
{E\over\sqrt{A}}=
{E_1\over\sqrt{A_1}}+
{E_2\over\sqrt{A_2}},
$$
so that $E>E_1+E_2$. The difference can be interpreted as an interaction
or binding energy, showing that it is energetically favourable
for separate concentrations of energy to coalesce.
Secondly, the term in the Ricci scalar may be integrated
by the Gauss-Bonnet theorem
(\eg Spivak 1979),
$$
\I\mu\R=8\pi(1-g),
$$
where the topological invariant $g$ is the genus (number of handles) of $S$:
$g=0$ for a sphere, $g=1$ for a torus, \etc.
Thirdly, $E$ reduces to the Hawking mass $M_H$
in the shear-free, twist-free case.
Finally, the shear term in $E$ is precisely the addition to the Hawking mass
required to yield zero in flat spacetime, as follows.

\ctitle{4. Flat spacetime}
Recall first the contracted Gauss equation (\eg Spivak 1979),
$$
\R+\t\tt-\half\sst=h^{ac}h^{bd}R_{abcd},
$$
which is a purely geometrical equation describing the embedding of $S$,
with $R_{abcd}$ being the Riemann tensor of $g_{ab}$.
\ms
In flat spacetime, commutativity of the null normals means that
the twist vanishes, $\o_a=0$,
and the Gauss equation becomes
$$
\R+\t\tt-\half\sst=0,
$$
so that $\H$ and $E$ vanish.
\ms
Alternatively, note that
the expansions and shears can be taken as equal and opposite,
$\t=-\tt$, $\s_{ab}=-\st_{ab}$,
by fixing the boost freedom of the normals.
For a 2-surface which lies entirely in one Euclidean 3-plane,
Euclidean surface theory (\eg Spivak 1979) can be applied.
Here, the Ricci scalar is twice the Gaussian curvature,
and so is defined by $\R=2\k_+\k_-$,
where the principal curvatures $\k_\pm$
are the roots of the eigenvalue equation
$$
\det\left(\root(\s_{ab}+\t h_{ab})-\k h_{ab}\right)=0,
$$
since the null second fundamental form is $\sqrt2$ times the
Euclidean second fundamental form.
This yields
$$
\sqrt2\k_\pm=\t\pm\sqrt{\half\sss},
$$
and so
$$
\R-\t^2+\half\sss=0,
$$
so that $\H$ and $E$ vanish.
\ms
Since $\sst\le0$, it also follows that the Hawking mass is non-positive
in flat spacetime, $M_H\le0$, vanishing only for shear-free surfaces.
This problem with the Hawking mass was well-known,
and its solution is simple: add the appropriate term in the shear,
as determined by the Gauss equation.
A similar argument shows that the mass of Geroch (1973)
is also non-positive in flat spacetime.
Horowitz and Schmidt (1982) showed that the Geroch mass
is always less than the Hawking mass.

\ctitle{5. Null infinity}
Asymptotically flat spacetimes were first studied
in the axisymmetric case by \Bondi (1962),
and in general by Sachs (1962).
They defined an asymptotic mass $M_B$ on slices of null infinity,
and showed that it decreased to the future,
which is interpreted as a loss of mass due to gravitational radiation.
Briefly sketching the approach:
on a foliation of null surfaces labelled by $u$,
spherical polar coordinates $(r,\tv,\pv)$ were introduced,
such that each surface $S$ of constant $(u,r)$ had area form
$\mu=r^2\a$, and hence area $A=4\pi r^2$.
Asymptotic expansions in negative powers of $r$ were assumed,
and the Bondi mass $M_B$ was defined as the integral over
$\a$ of a certain coefficient,
giving a generalisation of the Schwarzschild mass.
In the Appendix it is shown that the Bondi mass may be written as
$$
M_B=\l{1\over{8\pi}}\A\I\mu\C,
$$
where $C_{abcd}$ is the Weyl tensor of $g_{ab}$.
Thus the Bondi mass may be described as the average asymptotic Coulomb part
of the gravitational field.
The advantage of such an invariant expression in the present context
is that the radius $r$
does not appear explicitly except as the limit parameter,
so that the limit could simply be removed to yield a quasi-local mass
corresponding to $M_B$.
Using the vacuum Gauss equation
$$
\R+\t\tt-\half\sst=\C,
$$
the Bondi mass may be rewritten in terms of the fundamental forms as
$$
M_B=\l{1\over{8\pi}}\A\I\mu(\R+\t\tt-\half\sst).
$$
Comparing with the quasi-local energy $E$,
the asymptotic behaviour of the various terms may be translated from
the spin-coefficient expressions (\eg \S9.8 of Penrose and Rindler 1986) as
$$
\R=2r^{-2}+O(r^{-3}),\qquad
\t\tt=-2r^{-2}+O(r^{-3}),\qquad
\sst=O(r^{-3}),\qquad
\oo=O(r^{-6}),
$$
so that the asymptotic limit of $E$ exists, $E=O(1)$,
and equals the Bondi mass,
$$
M_B=\l E,
$$
since the twist term tends to zero.
Thus $E$ satisfies the criterion of
tending to the Bondi mass asymptotically, if the latter exists.
\ms
Note that the expression for Bondi-Sachs mass given by
Penrose (1963, 1964, 1967)
is an asymptotic limit of the Hawking mass,
$$
M_B=\lim_{\r\to\infty}M_H,
$$
which appears to differ from the previous expression by a term in the shears.
This is due to a different limit being taken,
using a foliation based on an affine parameter $\r=r+O(r^{-1})$,
as is explained in detail in the Appendix.
The affine expression for the Bondi mass
is used in most work on asymptotic flatness
(\eg Newman and Tod 1980, Winicour 1980, Wald 1984,
Penrose and Rindler 1986, Christodoulou and Klainerman 1989, Stewart 1990).
Other expressions are given by Bramson (1975) and Streubel (1978).
See also Winicour and Tamburino (1965),
Tamburino and Winicour (1966), Geroch (1977),
Geroch and Winicour (1981) and Walker (1982)
for the `linkage' formulation of asymptotic mass and angular momentum.
\ms
The positive-mass proofs of Ludvigsen and Vickers (1982),
Horowitz and Perry (1982) and Reula and Tod (1984)
all use the affine version,
as can be seen by comparing their expressions for mass
with the definitions of
Ludvigsen and Vickers (1981) and Horowitz and Tod (1982).
A positive-mass proof
has been given by Schoen and Yau (1982) using the original approach.

\ctitle{6. Spatial infinity}
Arnowitt \etal (1962) defined an asymptotic mass $M_{ADM}$
at spatial infinity, which is interpreted as the total mass of
an asymptotically flat spacetime.
The definition may be written in a more coordinate-independent way
(Ashtekar and Hansen 1978, Ashtekar 1980, Ashtekar and Horowitz 1982) as
$$
M_{ADM}=\lim_{\rho\to\infty}{1\over{8\pi}}\A\I\mu\C,
$$
with $\rho$ denoting the area radius of a family of spheres $S$
which approach spatial infinity in a spatial 3-surface (Appendix).
As at null infinity, the twist term in $E$ disappears in the limit,
so that
$$
M_{ADM}=\lim_{\rho\to\infty}E.
$$
Thus $M_B$ and $M_{ADM}$ are null and spatial limits of the same quantity
$E$, or equivalently of the average asymptotic Coulomb part of
the gravitational field.
This agreement has apparently not been noticed previously.
Indeed, in relating $M_{ADM}$ to the limit of $M_B$,
Ashtekar and Magnon-Ashtekar (1979)
found it necessary to require the vanishing at spatial infinity of
the shear term that apparently distinguishes
the area-radius definition from the affine version
(Ashtekar and Horowitz 1982).
Note that it is not clear whether $M_{ADM}$ is the limit of
the Bondi mass $M_B$ at spatial infinity,
since the limits are from different directions and may not coincide.
The situation is simply that if $M_B$ or $M_{ADM}$ exist,
then they are the appropriate limits of $E$.

\ctitle{7. On uniqueness}
The results of \S4--6 indicate that the freedom to choose
a sensible quasi-local energy, even on purely geometrical grounds,
is quite limited.
More precisely, assume that the energy can depend only on
the fundamental forms of the 2-surface,
including invariance under interchange and boosts
between the null normals.
Assume also that the energy is second-degree in $g_{ab}$,
\ie a linear combination of second derivatives and
quadratic first derivatives (\cf the discussion in \S11.2 of Wald 1984).
Then the only such functions are linear combinations of
$\R$, $\t\tt$, $\sst$ and $\oo$.
The coefficients of $\R$ and $\t\tt$ are determined by
consideration of metric spheres in flat spacetime
and in the Schwarzschild solution,
and the coefficient of $\sst$ is determined by
demanding that the energy vanish in flat spacetime, as in \S4.
This automatically yields the Bondi mass $M_B$ at null infinity,
as in \S5, and the \ADM mass at spatial infinity, as in \S6.
This leaves only the coefficient of $\oo$,
which is not so clearly determined,
but can be justified as a contribution due to angular momentum (Hayward 1993c).
The above assumptions are hardly compulsory but seem quite plausible,
and it is sensible at least
to exhaust such possibilites before resorting to more radical suggestions.
It is certainly remarkable that $E$ automatically satisfies these desiderata.

\ctitle{8. Examples}
In calculating the energy for particular spacetimes,
it is often most convenient to use the full Hamiltonian.
Alternatively,
in simple cases a transformation may be sought in which the line-element,
evaluated at the 2-surface $S$ in question, takes the form
$$
ds^2=-2d\xi\,d\eta+h_{ab}dx^adx^b,
$$
where $(x^1,x^2)$ are coordinates on $S$
and $(\xi,\eta)$ are affine parameters;
this is referred to subsequently as the standard form.
The fundamental forms may then be calculated by noting that
$\L_u=\p/\p\xi$ and $\L_v=\p/\p\eta$ in the basis used.
Also useful are the expressions $\mu\t=\mu_\xi$ and $\mu\tt=\mu_\eta$.
\ms
In the particular case of spherical symmetry,
a sphere of symmetry has zero twist and shears,
and the same results are found as for the Penrose mass according to Tod (1983)
or the Hawking mass.
Here spherical polar coordinates $(r,\tv,\pv)$ may be taken such that
$\mu=r^2\a$, and conseqently $A=4\pi r^2$,
$\R=2r^{-2}$, $\t=2r^{-1}r_\xi$, $\tt=2r^{-1}r_\eta$,
and so $$E=r(r_\xi r_\eta+\half).$$
This definition of gravitational energy
is often regarded as standard in spherical symmetry,
and was originally given in a different form by Misner and Sharp (1964).
A few examples follow, with interpretive comments.
\ms
{\it The Schwarzschild black-hole solution}
$$
ds^2=-\Sch dt^2+\Sch^{-1}dr^2+r^2\sph
$$
can be put in standard form by taking
$$
r_\xi=-r_\eta=-{1\over{\sqrt2}}\Sch^{1/2},\qquad
t_\xi=t_\eta={1\over{\sqrt2}}\Sch^{-1/2},
$$
and so $E=m$.
It should be emphasised that an interior solution matched to
the Schwarzschild exterior need not have total material mass $m$
as measured on a spatial 3-surface.
In this sense $E$ is not sensitive to the distribution of matter inside $S$,
being only a single number,
but rather measures the {\it effective} active gravitational mass
as felt on $S$.
Another way to see the necessity of considering the effective mass
is to note that the maximally extended Schwarzschild solution
has zero material mass, being a vacuum solution.
\ms
{\it The Reissner-Nordstr\"om charged black-hole solution}
$$
ds^2=-\RN dt^2+\RN^{-1}dr^2+r^2\sph
$$
can be put in standard form by taking
$$
r_\xi=-r_\eta=-{1\over{\sqrt2}}\RN^{1/2},\qquad
t_\xi=t_\eta={1\over{\sqrt2}}\RN^{-1/2},
$$
and so $E=m-\half e^2r^{-1}$.
The same result was obtained for the Penrose mass by Tod (1983),
who noted that the term in $e$ agrees with the linearised limit.
If this is regarded as a correction to $m$ which yields the effective mass,
then there is the interesting result that the field becomes repulsive
close to the $r=0$ singularity, a result which is also indicated by
the behaviour of geodesics (\S5.5 of Hawking and Ellis 1973).
For a non-extreme black hole, $e^2<m^2$,
such negative $E$ occurs inside the inner horizon,
which is widely regarded as unstable due to an `infinite blue-shift' effect,
in which case the inner region would not exist in practice.
Nevertheless, the example indicates that
a negative energy could be interpreted in a quasi-Newtonian way as
a net repulsion of the 2-surface (\cf Kulkarni \etal 1988).
\ms
{\it The Robertson-Walker cosmological solutions} are given by
$$
ds^2=-dt^2+a^2\bigl(d\psi^2+f^2\sph\bigr),
$$
where $a(t)$ and
$$
f(\psi)=
\cases{\sin\psi&for $k=1$,\cr\psi&for $k=0$,\cr\sinh\psi&for $k=-1$,\cr}
$$
where $k$ labels the spherical, flat and hyperbolic cases respectively.
This may be put in standard form with radius $r=af$ by taking
$$
\psi_\xi=-\psi_\eta={1\over{a\sqrt2}},\qquad
t_\xi=t_\eta={1\over{\sqrt2}}.
$$
The field equation
$a_t^2={8\over3}\pi a^2\rho-k$,
where $\rho$ is the density,
then yields $E={4\over3}\pi r^3\rho$. Recalling that $A=4\pi r^2$,
$E$ is just the product of density and the `area volume'
associated with the area radius $r$,
rather than the actual volume of a 3-surface inside $S$.
Again, this leads to the interpretation of $E$ as
the effective mass felt at $S$.
Note that $E$ is independent of the pressure,
which could for instance be negative,
as in the de Sitter and anti de Sitter solutions.
\ms
Although quasi-local mass is usually considered only for spheres,
the definition of $E$ applies to any compact orientable 2-surface,
and it is interesting to compare the results obtained for a flat torus
in a plane-symmetric spacetime with topological identifications.
The general plane-symmetric line-element is given by Szekeres (1972) as
$$
ds^2=-2e^{-K}du\,dv+e^{-P}(e^Q\cosh W dx^2-2\sinh W dx\,dy+e^{-Q}\cosh W dy^2),
$$
where $(K,P,Q,W)$ are functions of $(u,v)$,
and the toroidal identifications are
$$
(x,y)=(x+x_0,y+y_0).
$$
The metric may be put in standard form by taking
$$
u_\xi=v_\eta=e^{K/2},\qquad
u_\eta=v_\xi=0,
$$
and it is straightforward to calculate $\mu=e^{-P}dx\wedge dy$,
$A=e^{-P}x_0y_0$, $\R=0$, $\t\tt=e^KP_uP_v$,
$\sst=2e^K(Q_uQ_v\cosh^2W+W_uW_v)$, $\oo=0$, and hence
$$
E={1\over4}\left({x_0y_0\over{4\pi}}\right)^{3/2}e^{K-3P/2}
(P_uP_v-Q_uQ_v\cosh^2W-W_uW_v).
$$
\ms
{\it Plane wave spacetimes} are given by the above metric
depending only on $u$, or only on $v$, so that $E=0$.
This result is not surprising if
$E$ is interpreted as a measure of non-linear gravitational energy,
since plane waves propagate linearly.
\ms
{\it Colliding plane wave spacetimes},
introduced by Szekeres (1970, 1972)
and reviewed by Griffiths (1991),
have non-zero $E$, with no definite sign.
As an example, the solution of Khan and Penrose (1971),
describing the collision of two impulsive gravitational waves,
has $E<0$ after the collision, with $E$ becoming unbounded
at the singularity formed by the collision.
(The exact expression is long and unilluminating.)
The usual interpretation is that the two incoming waves
interact non-linearly, producing scattered radiation $\P$
whose magnitude grows without bound,
a property dependent on the plane symmetry.
The energy $E$ provides a measure of the non-linear interaction,
being zero before the collision and becoming unbounded at the singularity.
This illustrates both that $E$ is not a preserved quantity,
and that it is sensitive to gravitational interactions in vacuum.
Similar results are obtained for generic colliding plane waves,
and for asymptotically plane waves (Hayward 1990, 1992):
$E$ equals $\Re\P$ up to a factor,
with $\P$ generically becoming unbounded after the collision.

\ctitle{9. Energy inequalities}
On the question of whether a quasi-local energy
should be manifestly non-negative, it should be noted that
negative-energy matter can be described quite consistently
in general relativity, and in such a situation it would be preferable
for the quasi-local energy also to become negative.
The examples of \S8 show that $E$ is positive
for certain physically familiar gravitational fields,
but that negative $E$ may occur in other circumstances, even in vacuum.
{}From purely quasi-local arguments,
the fundamental forms are freely specifiable on any one 2-surface,
and so it is easy to construct 2-surfaces with negative $E$.
Nevertheless, it may be possible to find a positivity theorem
based on global assumptions, such as asymptotic flatness,
global hyperbolicity \etc,
together with a local energy condition on the matter.
A weaker possibility would be positivity for spacetimes sufficiently close
to flat spacetime (Christodoulou and Klainerman 1989),
or for sufficiently round spheres (\eg Christodoulou and Yau 1988).
Conversely,
$E$ may be a suitable quantity to control the gravitational field
in the context of global existence theorems or cosmic censorship.
In this context it is interesting to note that in the 2-dimensional
dilaton gravity theory of Callan \etal (1992),
positive gravitational energy is associated with trapped spatial singularities,
and negative energy with naked singularities (Hayward 1993b).
In spherical symmetry,
there is a similar relationship between the sign of $E$
and the signature and trapping of singularities (\cf Christodoulou 1991).
\ms
A related question is that of monotonicity,
\eg whether $E$ increases as $A$ increases along a foliation of 2-surfaces.
Here it may be useful to consider {\it well-oriented} surfaces such that
$\t\tt<0$,
which have an orientation determined according to which expansion
is negative and which positive.
For such surfaces, the Hawking mass decreases internally
and increases externally (Eardley 1979), which can be interpreted as
a quasi-local generalisation of the Bondi mass-loss result.
A similar result for $E$ would be of interest,
but seems more difficult to obtain.
\ms
Similar comments apply to energy inequalities such as
the isoperimetric inequality (Gibbons 1972, Penrose 1973)
and the hoop conjecture (Thorne 1972) for trapped surfaces.
For a marginally trapped sphere, $\t\tt=0$ and $g=0$,
$$
\sqrt{{16\pi E^2\over A}}=1-{1\over{8\pi}}\I\mu(\half\sst+2\oo),
$$
and consequently the isoperimetric inequality $16\pi E^2\ge A$
can be violated by a suitable quasi-local choice of shears and twist.
Again, global assumptions may be relevant.
Note however that the twist and internal shear vanish on a Killing horizon
(Tod 1986),
so that a {\it static} black hole satisfies the isoperimetric {\it equality}
$16\pi E^2=A$ on the horizon, a more precise result than hitherto obtained
(Ludvigsen and Vickers 1983b, Tod 1985, 1986, 1992).
The very triviality of this result suggests that
$E$ may be an appropriate definition for such purposes.
Expressed equivalently, $E$ is the irreducible mass on a static horizon,
\ie the mass of a Schwarzschild black hole with the same area
(Christodoulou and Ruffini 1971).

\ctitle{10. Criteria for quasi-local energy}
As noted in the Introduction, there is now a plethora of
suggested quasi-local energies, which disagree even in
the simple Reissner-Nordstr\"om and Kerr cases (Bergqvist 1992a).
The ambiguity can only be resolved by
settling on generally agreed criteria,
and I would like to make a few such suggestions.
%and to encourage further thought in this direction.
Most fundamentally, the energy should be {\it well-defined}
on any compact orientable spatial 2-surface,
or at least for the zero genus case.
This firm but fair requirement disqualifies almost all of the contenders,
which depend on particular symmetries
(Komar 1959, Katz \etal 1988, Kulkarni \etal 1988,
Bergqvist and Ludvigsen 1989),
a preferred 3-surface (Geroch 1973, Nester 1991, Brown and York 1993),
global requirements (Ludvigsen and Vickers 1983a, Bartnik 1989)
or the existence of solutions to particular equations
(Penrose 1982, Dougan and Mason 1991, Bergqvist 1992b).
To the best of my knowledge,
this leaves only the Hawking mass as an unambiguous quasi-local definition.
\ms
The other basic criteria are that the energy vanishes in flat spacetime,
and yields the Bondi mass at null infinity
and the \ADM mass at spatial infinity.
It would also be of interest to investigate
the Newtonian and linearised limits.
Eardley (1979) and Christodoulou and Yau (1988)
give a comparable list of criteria,
of which positivity and monotonicity remain as open questions for $E$.
Bergqvist (1992b) also gives a list of criteria,
and shows that there are infinitely many possibilities which satisfy it.
Such extravagance can be curtailed by demanding
that the definition should depend on the fundamental forms only,
as these are the only purely geometrical quantities
associated with a 2-surface.
The fundamental forms are also the dynamically independent
parts of the gravitational field on a 2-surface,
or equivalently the free gravitational data.
\ms
Even with such criteria,
the arguments of \S7 show that a unique `correct' quasi-local energy
will require further (presumably dynamical) justification.
Some of the suggested quasi-local energies involve spurious dynamics,
in the sense that they are based on the introduction of fields
satisfying various particular equations
other than the Einstein equations.
The main point of this article is that the `2+2' Hamiltonian $\H$
is a dynamically preferred quantity which can justifiably be interpreted
as the energy density of the Einstein gravitational field
referred to a 2-surface element.
The fact that the associated quasi-local energy $E$
has the agreeable properties described herein
may be taken as firm supporting evidence. In view of this,
are there any objections to using the `2+2' Hamiltonian of a field
to {\it define} its quasi-local energy?
Is $E$ the long-sought gravitational energy?
\vfill
\ni{\bf Acknowledgements.}\nl
I would like to thank J\"urgen Ehlers, J\"org Frauendiener,
Helmut Friedrich, Alan Rendall,
Wolfgang Rindler, Bernd Schmidt and John Stewart
for various relevant information, suggestions and questions.
This research was supported by the European Science Exchange Programme.
\eject
\ctitle{Appendix: the Bondi mass}
When defining the Bondi mass as a limit of a quasi-local energy
along a null foliation of 2-surfaces $S$,
there is an important subtlety,
namely that the limit depends on the foliation used,
and in particular whether the foliation is based on
an affine parameter or a luminosity parameter (area radius).
Conversely, different quasi-local integrals must be used
in order to obtain the same limit.\footnote\dag
{I am grateful to J\"org Frauendiener and John Stewart for clarifying
an earlier confusion on this point.}
The following is based on a calculation of J\"org Frauendiener
(private communication).
\ms
Assume a parameter $s$ which is related to the area radius $r$ by
$$r=s+\half fs^{-1}+O(s^{-2})$$ for some $f$.
Take a spin-basis $(l,n,m,\m)$ such that $m$ spans $S$ and $l=\p/\p s$.
The Newman-Penrose convergence $\rho$ is then given by
$\rho=-s^{-1}+fs^{-3}+O(s^{-4})$.
\ms
The original definition of the Bondi mass (\Bondi 1962, Sachs 1962)
involved the Bondi coordinates $(u,r,\tv,\pv)$,
where $u$ is null and the area 2-form of a constant-$(u,r)$ surface is
$\mu=r^2\sin\tv\,d\tv\wedge d\pv$.
Namely, the definition is
$$M_B={1\over{4\pi}}\int M\a,$$
where the mass aspect $M$ is defined by the expansion
$${g^{rr}\over{g^{ru}}}=-1+2Mr^{-1}+O(r^{-2}),$$
or equivalently by
$$M=\half\l r\left({g^{rr}\over{g^{ru}}}+1\right).$$
This may be translated in terms of the parameter $s$ by
$$g^{rr}=g^{ss}(1-\half fs^{-2})^2+g^{su}f_us^{-1}+O(s^{-2}),\qquad
g^{ru}=g^{su}(1-\half fs^{-2})+O(s^{-3}).$$
Consider two special choices of $s$.
Firstly, take $s$ to be the affine parameter $\r$
used in \S9.8 of Penrose and Rindler (1986), for which
$$f=-\sh^0\sb^0,\qquad
g^{\r u}=1,\qquad
g^{\r\r}=-1-2\Re\Ph^0\r^{-1}+O(\r^{-2}),$$
where $\Re$ denotes the real part,
$$\sh=\sh^0\r^{-2}+O(\r^{-3}),\qquad
\Ph=\Ph^0\r^{-3}+O(\r^{-4}),$$
and $\sh$ and $\Ph$ are the Newman-Penrose shear and complex `Coulomb' term
respectively.
Hence the mass aspect is given by
$$M=-\Re\Ph^0+\half f_u=-\Re(\Ph^0+\sh^0\sb^0_u)=-\Re(\Ph^0-\sh^0\sh^0{}').$$
Alternatively, taking $s$ to be the area radius $r$,
$$f=0,\qquad
{g^{rr}\over{g^{ru}}}=-1-2\Re\P^0r^{-1}+O(r^{-2}),$$
and so simply
$$M=-\Re\P^0.$$
There are various equivalent expressions
(\eg Ashtekar and Magnon-Ashtekar 1979), since
$$\C=2e^{-m}h^{ac}l^bn^dC_{abcd}=-2e^{-2m}l^an^bl^cn^dC_{abcd}=-4\Re\P,$$
using the null normals $l^a$ and $n^a$ (\S2).
\ms
The above calculation shows that
the `Coulomb' terms for the spin-bases adapted to $r$ and $\r$
are related by $$\Re\P^0=\Re(\Ph^0-\sh^0\sh^0{}'),$$
essentially because $$r=\r-\half\sh^0\sb^0\r^{-1}+O(\r^{-2}).$$
Consequently, the Bondi mass can be expressed in either of the forms
$$M_B=-\l{1\over{2\pi}}\A\I\mu\,\Re\P
=-\lh{1\over{2\pi}}\Ah\Ih\mh\,\Re(\Ph-\sh\sh').$$
These are limits of
the quasi-local energy $E$ and the Hawking mass $M_H$ respectively,
$$M_B=\l E(S)=\lh M_H(\Sh).$$
Thus the Bondi mass is the limit of the Hawking mass
along the affine foliation used by Penrose and Rindler (1986),
and is also the limit of the quasi-local energy $E$
along a foliation parametrised by area radius.
The area radius or luminosity parameter was the original choice (\Bondi 1962)
and an affine parameter is used in most other work
(\eg Newman and Unti 1962, Newman and Penrose 1968,
Dixon 1970, Newman and Tod 1980,
Penrose and Rindler 1986, Stewart 1990).
The area radius is in many ways the most natural choice,
for instance from purely geometrical considerations,
or from the link with the \ADM mass (\S6).
Additionally,
note that other affine parameters would also require different expressions,
with the particular affine parameter $\r$
being chosen to agree with the area radius to the highest possible order.
\np
\begingroup
\parindent=0pt\everypar={\global\hangindent=20pt\hangafter=1}
{\bf References}\par
\ref{Arnowitt R, Deser S \& Misner C W 1962 in}
{Gravitation, an Introduction to Current Research}\ns{ed Witten (Wiley)}
\ref{Ashtekar A 1980 in}{General Relativity and Gravitation}
\ns{ed Held (Plenum)}
\ref{Ashtekar A \& Hansen R O 1978}\JMP{19}{1542}
\ref{Ashtekar A \& Horowitz G T 1982}\PL{89A}{181}
\ref{Ashtekar A \& Magnon-Ashtekar A 1979}\PRL{43}{181}
\ref{Bartnik R 1989}\PRL{62}{2346}
\ref{Bergqvist G 1992a}\CQG9{1753}
\ref{Bergqvist G 1992b}\CQG9{1917}
\ref{Bergqvist G \& Ludvigsen M 1989}\CQG6{L133}
\ref{Bondi H, van der Burg M G J \& Metzner A W K 1962}\PRS{A269}{21}
\ref{Bramson B D 1975}\PRS{A341}{463}
\ref{Brown J D \& York J W 1993}\PR{D47}{1407}
\ref{Callan C G, Giddings S B, Harvey J A \& Strominger A 1992}\PR{D45}{R1005}
\ref{Christodoulou D 1991}\CPAM{44}{339}
\ref{Christodoulou D \& Klainerman S 1989}
{The global nonlinear stability of the Minkowski space}\ns{(preprint)}
\ref{Christodoulou D \& Ruffini R 1971}\PR{D4}{3552}
\ref{Christodoulou D \& Yau S-T 1988 in}
{Mathematics and General Relativity}
\ns{ed Isenberg (American Mathematical Society)}
\ref{Dixon W G 1970}\JMP{11}{1238}
\ref{Dougan A J \& Mason L J 1991}\PRL{67}{2119}
\ref{Eardley D M 1979 in}{Sources of Gravitational Radiation}
\ns{ed Smarr (Cambridge University Press)}
\ref{Geroch R P 1973}\ANY{224}{108}
\ref{Geroch R P 1977 in}{Asymptotic Structure of Space-Time}
\ns{ed Esposito \& Witten (Plenum)}
\ref{Geroch R \& Winicour J 1981}\JMP{22}{803}
\ref{Gibbons G W 1972}\CMP{27}{87}
\ref{Griffiths J B 1991}{Colliding Waves in General Relativity}
\ns{(Oxford University Press)}
\ref{Hawking S W 1968}\JMP9{598}
\ref{Hawking S W \& Ellis G F R 1973}{The Large Scale Structure of Space-Time}
\ns{(Cambridge University Press)}
\ref{Hayward S A 1990}\CQG7{1117}
\ref{Hayward S A 1992}{Asymptotically plane Einstein-Klein-Gordon waves}\ns
{(preprint)}
\ref{Hayward S A 1993a}{Dual-null dynamics of the Einstein field,
\CQG}{10}{(in press)}
\ref{Hayward S A 1993b}{Cosmic censorship in two-dimensional dilaton gravity,
\CQG}{10}{(in press)}
\ref{Hayward S A 1993c}{General laws of black-hole dynamics}\ns
{(gr-qc/9303006)}
\ref{Helfer A D 1992}\CQG9{1001}
\ref{Horowitz G T \& Tod P 1982}\CMP{85}{429}
\ref{Horowitz G T \& Perry M J 1982}\PRL{48}{371}
\ref{Horowitz G T \& Schmidt B 1982}\PRS{A381}{215}
\ref{Katz J, Lynden-Bell D \& Israel W 1988}\CQG5{971}
\ref{Khan K A \& Penrose R 1971}{Nature}{229}{185}
\ref{Komar A 1959}\PR{113}{934}
\ref{Kulkarni R, Chellathurai V \& Dadhich N 1988}\CQG5{1443}
\ref{Ludvigsen M \& Vickers J A G 1981}\JP{A14}{L389}
\ref{Ludvigsen M \& Vickers J A G 1982}\JP{A15}{L67}
\ref{Ludvigsen M \& Vickers J A G 1983a}\JP{A16}{1155}
\ref{Ludvigsen M \& Vickers J A G 1983b}\JP{A16}{3349}
\ref{Misner C W \& Sharp D H 1964}\PR{136B}{571}
\ref{Nester J M 1991}\CQG8{L19}
\ref{Newman E T \& Penrose R 1962}\JMP3{566}
\ref{Newman E T \& Penrose R 1968}\PRS{A305}{175}
\ref{Newman E T \& Tod K P 1980 in}{General Relativity and Gravitation}
\ns{ed Held (Plenum)}
\ref{Newman E T \& Unti T W J 1962}\JMP3{891}
\ref{Penrose R 1963}\PRL{10}{66}
\ref{Penrose R 1964 in}{Relativity, Groups and Topology}
\ns{ed de Witt \& de Witt (Gordon \& Breach)}
\ref{Penrose R 1967 in}{Relativity Theory and Astrophysics Vol.\ 1}
\ns{ed Ehlers (American Mathematical Society)}
\ref{Penrose R 1973}\ANY{224}{125}
\ref{Penrose R 1982}\PRS{A381}{53}
\ref{Penrose R \& Rindler W 1984}{Spinors and Space-Time Vol.\ 1}
\ns{(Cambridge University Press)}
\ref{Penrose R \& Rindler W 1986}{Spinors and Space-Time Vol.\ 2}
\ns{(Cambridge University Press)}
\ref{Reula O \& Tod K P 1984}\JMP{25}{1004}
\ref{Schoen R \& Yau S-T 1982}\PRL{48}{369}
\ref{Sachs R K 1962}\PRS{A270}{103}
\ref{Spivak M 1979}
{A Comprehensive Introduction to Differential Geometry}
\ns{(Publish or Perish)}
\ref{Stewart J 1990}{Advanced General Relativity}
\ns{(Cambridge University Press)}
\ref{Streubel M 1978}\GRG9{551}
\ref{Szekeres P 1970}{Nature}{228}{1183}
\ref{Szekeres P 1972}\JMP{13}{286}
\ref{Tamburino L A \& Winicour J H 1966}\PR{150}{1039}
\ref{Thorne K S 1972 in}{Magic without Magic}\ns{ed Klauder (Freeman)}
\ref{Tod K P 1983}\PRS{A388}{457}
\ref{Tod K P 1985}\CQG2{L65}
\ref{Tod K P 1986}\CQG3{1169}
\ref{Tod K P 1992}\CQG9{1581}
\ref{Wald R M 1984}{General Relativity}\ns{(University of Chicago Press)}
\ref{Walker M 1982 in}{Gravitational Radiation}
\ns{ed Deruelle \& Piran (North-Holland)}
\ref{Winicour J 1968}\JMP9{861}
\ref{Winicour J 1980 in}{General Relativity and Gravitation}
\ns{ed Held (Plenum)}
\ref{Winicour J \& Tamburino L 1965}\PRL{15}{601}
\endgroup
\bye